\documentclass[onecolumn,preprintnumbers,amsmath,amssymb]{revtex4}
\usepackage{graphicx}
\usepackage{dcolumn}
\usepackage{bm}

\usepackage{amsthm}
\usepackage{epsfig}
\usepackage{subfigure}
\usepackage{amsmath}
\usepackage{indentfirst}
\usepackage{appendix}

\selectfont

\begin{document}

\preprint{}
\title{High-order rogue waves of a long wave-short wave model }
\author{Junchao Chen$^{1}$\footnote{email:junchaochen@aliyun.com}, Liangyuan Chen$^{2}$, Bao-Feng Feng$^{3}$, Ken-ichi Maruno$^{4}$}
\affiliation{ $^{1}$Department of Mathematics, Lishui University, Lishui, 323000, People's Republic of China}
\affiliation{ $^{2}$Department of Photoelectric Engineering, Lishui University, Lishui 323000, People's Republic of China}
\affiliation{ $^{3}$Department of Mathematics, The University of Texas-Rio Grande Valley, Edinburg, TX 78541, USA}
\affiliation{ $^{4}$Department of Applied Mathematics, School of Fundamental Science and Engineering,
Waseda University, 3-4-1 Okubo, Shinjuku-ku, Tokyo 169-8555, Japan}
\date{\today}

\begin{abstract}
The long wave-short wave model describes the interaction between the long wave and the short wave.
Exact higher-order rational solution expressed by determinants is calculated via the Hirota's bilinear method and the KP hierarchy reduction.
It is found that the fundamental rogue wave for the short wave can be classified into three different patterns: bright, intermediate and
dark ones, whereas the rogue wave for the long wave is always bright type.
The higher-order rogue waves correspond to the superposition of fundamental rogue waves.
The modulation instability  analysis show that the condition of the baseband modulation instability where an unstable continuous-wave background corresponds to perturbations with infinitesimally small frequencies, coincides with the condition for the existence of rogue-wave solutions.

PACS number(s): 05.45.Yv, 02.30.Ik, 02.30.Jr, 11.10.Lm
\end{abstract}

\maketitle

\section{Introduction}

Rogue waves (RWs) or freak waves are rare phenomena that the large amplitudes appear from the background with the instability and unpredictability.
Such extreme wave can be observed in various different contexts such as
oceanography \cite{kharif2009rogue}, hydrodynamic \cite{onorato2013rogue,chabchoub2011rogue}, Bose-Einstein condensate \cite{bludov2009matter}, plasma \cite{bailung2011observation} and nonlinear optic \cite{onorato2013rogue,solli2007optical,kibler2010peregrine}.
Mathematically, Peregrine soliton of the nonlinear Schr\"{o}dinger (NLS) equation serves as
a prototype of the RW, in which its structure is localized in temporal-spatial distribution plane and its maximum amplitude attains three times the background \cite{peregrine1983water}.
Since the higher-order RW was excited experimentally in wave tanks \cite{chabchoub2013observation,chabchoub2012observation}, a hierarchy of higher-order analytic RW solutions which indicate the superposition of elementary RW has been found in nonlinear integrable systems \cite{akhmediev2009rogue,kedziora2012second,dubard2010multi,guo2012nonlinear,ling2016multisoliton,ohta2012general,he2010generating,wen2015generalized,yang2015rogue,chen2015rational}.
Moreover, in contrast to the scalar system,
recent studies have shown that multicomponent coupled systems may allow
some novel patterns of RW such as dark and four-petaled types \cite{guo2011roguewave,ling2012highorder,baronio2012solutions,zhao2013roguewave,baronio2013solutions,zhao2016high,ling2016darboux,chen2014dark,chow2013rogue,mu2015dynamics,wang2015roguewave,zhang2017solitons,wen2015modulational}.

Modulation instability (MI) refers to the basic process
that the growth of perturbations emerges on an unstable continuous
wave background \cite{zakharov2009modulation}.
In the explanations of the generation mechanism for the RW,
MI has been found to be closely linked with
the RW excitation in nonlinear dispersive systems \cite{baronio2014vector,baronio2015baseband,zhao2016quantitative,biondini2016universal}.
It has been shown that RW modeled as rational solutions only exist in the
subset of parameters where MI is present if and only if the unstable sideband spectrum also contains continuous wave or zero-frequency perturbations as a limiting case \cite{baronio2014vector,baronio2015baseband}.

To discover how waves with different length scales (frequencies) interact and affect each other,
Benny established the general theory for the interaction between the long wave (LW) and the short wave (SW) \cite{benney1977general}.
Particularly, under the certain resonance condition, namely,
the phase velocity of the LW is equal to the group velocity of the SW,
energy exchange can be anticipated and the resonance interaction occurs \cite{benney1977general}.
Such resonance process may appear in a variety of physical settings such as
capillary-gravity waves, internal-surface waves, short and long gravity
waves on fluids of finite depth and the breakdown of laminar flow \cite{benney1977general,newell1978long,newell1979general}.

The aim of the present work is investigate RWs for a coupled system with the LW interacting the SW via the KP hierarchy reduction, and discuses the mechanism for the RW excitation through the MI anylsis.
The outline of the present paper is as follows.
In Sec. II, general analytical higher-order rational solutions in terms of determinants with algebraic elements are derived via the Hirota's bilinear method and the KP hierarchy reduction.
In Sec. III, local structures of RWs are analyzed and show that the SW possesses bright, intermediate and
dark patterns, whereas the LW always exhibits bright state in the fundamental RW.
The higher-order RWs indicate the superposition of fundamental ones and
interaction behaviors among different types of the RW can't occur in such pure higher-order rational solutions.
In Sec. IV, MI analysis is carried out to find that the condition of baseband MI coincides with the condition for the existence of rogue-wave solutions. Numerical simulations are also provided to show RW excitation in the regime of baseband MI.
A discussion is given and results are summarized in Sec. V.

\section{ High-order rational solution in the determinant form}

Based on the Benney theory for the interaction between the SW and the LW \cite{benney1977general}, an integrable long wave-short wave (LWSW) model \cite{newell1978long,newell1979general} is proposed
\begin{eqnarray}
\label{dyo-rw-1}&& {\rm i}S_t +S_{xx} - 2{\rm i} S_x L =0 ,\\
\label{dyo-rw-2} && L_t=-2\sigma (|S|^2)_x,
\end{eqnarray}
where $S=S(x,t)$ represents the envelope of the short wave and $L=L(x,t)$ denotes the
amplitude of the long wave.
The complete integrability of the LWSW model (\ref{dyo-rw-1})-(\ref{dyo-rw-2}) was tested by Painlev\'{e} analysis \cite{chowdhury1986complete}.
Ling et al. constructed its Darboux transformation and found a closed multi-soliton solution formula \cite{ling2011long,huang2013darboux}.
A class of cusp solution for the LWSW model (\ref{dyo-rw-1})-(\ref{dyo-rw-2}) was derived by using the dressing method \cite{zhu2015cusp}.
Geng et al. provided this coupled system's algebro-geometric constructions and their explicit theta function representations \cite{geng2017algebro}.

By the dependent variable transformation
\begin{eqnarray}\label{trans}
S=\rho {\rm e}^{{\rm i}[\alpha x- \alpha^2t] }\frac{g}{f^*},\ \
L={\rm i}\frac{\partial}{\partial x}\ln\frac{f^*}{f},
\end{eqnarray}
where $\rho$ and $\alpha$ are real parameters, the LWSW model (\ref{dyo-rw-1})-(\ref{dyo-rw-2}) can be cast into the bilinear form
\begin{eqnarray}
\label{bilinear-01}&& ({\rm i}D_t +2{\rm i}\alpha D_x +D^2_x)g \cdot f=0, \\
\label{bilinear-02}&& {\rm i} D_tf\cdot f^* = D^2_xf\cdot f^* ,\\
\label{bilinear-03}&& {\rm i} D_tf\cdot f^*=-2\sigma \rho^2(|f|^2-|g|^2).
\end{eqnarray}
where $f$ and $g$ are complex variables, $*$ denotes the complex conjugation and $D$ is
Hirota's bilinear differential operator.
Then we first present the general rational solutions of the LWSW model (\ref{dyo-rw-1})-(\ref{dyo-rw-2}) in the following theorem.
The proof of this theorem is given in the Appendix.

\textbf{Theorem 2.1}
The LWSW equations (\ref{dyo-rw-1})-(\ref{dyo-rw-2}) have the rational solutions (\ref{trans}) with the tau functions
$f$ and $g$ given by $N\times N$ determinants
\begin{eqnarray}\label{theoremeq01}
f=\tau_{-1,0},\ \ g=\tau_{-1,1}
\end{eqnarray}
where
\begin{eqnarray}\label{theoremeq02}
\tau_{n,k}=\det_{1\leq i,j\leq N}\left(m^{(N-i,N-j,n,k)}_{2i-1,2j-1}\right)
=\begin{vmatrix}
m_{11}^{(N-1,N-1,n,k)} & m_{13}^{(N-1,N-2,n,k)} &\cdots & m_{1,2N-1}^{(N-1,0,n,k)} \\
m_{31}^{(N-2,N-1,n,k)} & m_{33}^{(N-2,N-2,n,k)} &\cdots & m_{3,2N-1}^{(N-2,0,n,k)} \\
\vdots &\vdots & & \vdots \\
m_{2N-1,1}^{(0,N-1,n,k)} &m_{2N-1,3}^{(0,N-2,n,k)} & \cdots & m_{2N-1,2N-1}^{(0,0,n,k)}
\end{vmatrix},
\end{eqnarray}
and the matrix elements are defined by
\begin{eqnarray}\label{theoremeq03}
m^{(\nu,\mu,n,k)}_{i,j}= \left. \sum_{l=0}^i \sum_{s=0}^j  \frac{ a_l^{(\nu)} }{(i-l)!} \frac{ a_s^{(\mu)*} }{(j-s)!} [(p-{\rm i}\alpha)\partial_p]^{i-l} [(q+{\rm i}\alpha)\partial_q]^{j-s} m^{(n,k)} \right|_{p=\zeta,q=\zeta^*},
\end{eqnarray}
with
\begin{eqnarray}
\label{theoremeq04}&& m^{(n,k)}=\frac{{\rm i}p}{p+q} \left(-\frac{p}{q}\right)^n \left(-\frac{p-{\rm i}\alpha}{q+{\rm i}\alpha}\right)^k e^{(p+q)x -(p^2-q^2) {\rm i}t},\\
\label{theoremeq05}&& a^{(\nu+1)}_{l}=\sum^{l}_{j=0} \frac{2^{j+2} (p-{\rm i}\alpha)^2 + (-1)^{j}  \frac{2{\rm i}\sigma\alpha \rho^2}{p-{\rm i}\alpha} +2{\rm i}\alpha(p-{\rm i}\alpha)}{ (j+2)!} a^{(\nu)}_{l-j},
\ \ \nu=0,1,2,\cdots,
\end{eqnarray}
and $\zeta$ need to satisfies the relation
\begin{eqnarray}\label{theoremeq06}
2\zeta- \frac{2\sigma{\rm i}\alpha\rho^2}{(\zeta-{\rm i}\alpha)^2}=0.
\end{eqnarray}

\section{Dynamics of rogue wave solutions}

\subsection{Fundamental rogue wave}

According to Theorem 2.1, in order to obtain the first-order rogue wave, we need to take $N=1$ in Eqs.(\ref{theoremeq01})-(\ref{theoremeq06}).
For simplicity, we set $a^{(0)}_0=b^{(0)}_0=1$ and $a^{(0)}_1=b^{(0)}_1=0$, then the functions $f$ and $g$ take the form
\begin{eqnarray}
&& f=\frac{1}{\zeta^3_1} e^{2\zeta_1(x-2\zeta_2t)} F,\\
&& g=\frac{1}{\zeta^3_1} e^{2\zeta_1(x-2\zeta_2t)}
\left[- \frac{\zeta_1+{\rm i}(\zeta_2-\alpha)}{\zeta_1-{\rm i}(\zeta_2-\alpha)} \right]
\left[F + l_1 -{\rm i} l_2 \right] ,
\end{eqnarray}
with
\begin{eqnarray}
&& F=-\Delta \left[ \frac{\zeta_2}{2}\theta^2_1 +2\zeta_2\theta^2_2 +\theta_0 +{\rm i} \frac{\zeta_1  }{2} (h^2_1+h^2_2) \right],
\end{eqnarray}
and $\Delta = \zeta^2_1 +(\zeta_2-\alpha)^2$,
$\theta_1= \zeta_1 x -2\zeta_1\zeta_2 t -\frac{1}{2}$,
$\theta_2= \zeta^2_1 t -\frac{\zeta_1}{4\zeta_2}$,
$\theta_0=\frac{\zeta^2_2-\zeta^2_1}{4\zeta_2}$,
$h_1=\zeta_1(x-2\zeta_2t)$, $h_2=2\zeta^2_1 t$,
$l_1=\zeta^2_1 \left(\zeta_1l_0 + \frac{\zeta_2}{2}\right)$,
$l_2=\zeta_1\zeta_2\left( \zeta_1 l_0 +\frac{\zeta_2-\alpha}{2} \right)$,
 and $l_0=(\alpha-\zeta_2)x -2 (\zeta^2_1-\zeta^2_2+\alpha \zeta_2)t$.

Therefore, the fundamental rogue wave solution for the LWSW model (\ref{dyo-rw-1})-(\ref{dyo-rw-2}) reads
\begin{eqnarray}
&& S=\rho {\rm e}^{{\rm i}[\alpha x- \alpha^2t] } \left[- \frac{\zeta_1+{\rm i}(\zeta_2-\alpha)}{\zeta_1-{\rm i}(\zeta_2-\alpha)} \right]
\left\{1- \frac{\zeta^2_1 \left(\zeta_1l_0 + \frac{\zeta_2}{2}\right) -{\rm i} \left[\zeta_1\zeta_2\left( \zeta_1 l_0 +\frac{\zeta_2-\alpha}{2} \right)-\Delta \zeta_1   (h^2_1+h^2_2) \right] }{\Delta \left[ \frac{\zeta_2}{2}\theta^2_1 +2\zeta_2\theta^2_2 +\theta_0 -{\rm i} \frac{\zeta_1  }{2} (h^2_1+h^2_2) \right]} \right\} ,\\
&&
L= - 4\frac{\zeta^2_1[\zeta_2\theta_1 (h^2_1+h^2_2-h_1\theta_1) -2h_1 (\theta_0+2\zeta_2\theta^2_2) ]}{\left(\zeta_2\theta^2_1 +4\zeta_2\theta^2_2 +2\theta_0 \right)^2  +  \zeta^2_1 (h^2_1+h^2_2)^2  }.
\end{eqnarray}

Furthermore, the modular square of the SW component $|S|^2$ possesses extrema (turning points where the first derivatives vanish)
\begin{eqnarray}
&& (x_1,t_1)= \left(\frac{\zeta^2_2}{\zeta_1(\zeta^2_1+\zeta^2_2)}, \frac{1}{4}\frac{\zeta_2}{\zeta_1(\zeta^2_1+\zeta^2_2)} \right),\\
&&
(x_2,t_2)=\left( \frac{\zeta_2(\alpha^2-\zeta^2_1-\zeta^2_2)}{2\zeta_1\Delta_3}-\frac{2\Delta_5\mu_1}{\Delta_1\Delta_3\zeta^2_1}, \frac{\mu_1}{\Delta_1\zeta^2_1}  \right),
\\
&&
(x_3,t_3)=\left( -\frac{\zeta_2(\alpha^2-\zeta^2_1-\zeta^2_2)}{2\zeta_1\Delta_6}+\frac{2\Delta_7\mu_2}{\Delta_1\Delta_6\zeta_1}, \frac{\mu_2}{\Delta_1\zeta^2_1}  \right),
\end{eqnarray}
with
$ \mu_1= \frac{\zeta_1\zeta_2\Delta_1}{4(\zeta^2_1+\zeta^2_2)} \pm \frac{ \sqrt{-\zeta_2\Delta_2\Delta^2_3\Delta_4} }{4(\zeta^2_1+\zeta^2_2)\Delta_2}$,
$ \mu_2= \frac{\zeta_2\Delta_1}{4(\zeta^2_1+\zeta^2_2)} \pm \frac{ \sqrt{\zeta_2\Delta^2_6\Delta_8\Delta_9} }{4(\zeta^2_1+\zeta^2_2)\Delta_9}$,
$ \Delta_1= \zeta^2_1 +(\zeta_2-\alpha)^2$,
$\Delta_2= \zeta^2_1 +(\zeta_2+\alpha)^2$,
$\Delta_3=\alpha (2\zeta^2_1 + \alpha \zeta_2) - \zeta_2(\zeta^2_1+\zeta^2_2)$,
$\Delta_4= \zeta_2(\zeta^2_1+\zeta^2_2) - 2(2\zeta^2_1 + \zeta^2_2)\alpha +\zeta_2\alpha^2$,
$\Delta_5=(\zeta^2_2-\zeta^2_1)(\zeta^2_1+\zeta^2_2-\alpha^2)-4\alpha\zeta_2\zeta^2_1$,
$
\Delta_6=3\zeta_2(\zeta^2_1+\zeta^2_2)-2\alpha (\zeta^2_1+2\zeta^2_2) +\zeta_2\alpha^2$,
$
\Delta_7=-(\zeta^2_1+\zeta^2_2)(\zeta^2_1-5\zeta^2_2)-4\zeta_2(\zeta^2_1+2\zeta^2_2)\alpha +(\zeta^2_1+3\zeta^2_2)\alpha^2$,
$
\Delta_8= 3\zeta_2(\zeta^2_1+\zeta^2_2) -2(2\zeta^2_1+3\zeta^2_2)\alpha+3\zeta_2\alpha^2$
and
$ \Delta_9=(\zeta^2_1+\zeta^2_2)(\zeta^2_1+4\zeta^2_2)-2\zeta_2(3\zeta^2_1+4\zeta^2_2)\alpha +(\zeta^2_1+4\zeta^2_2)\alpha^2$.

Note that $(x_2, t_2)$ are also two characteristic points, at which the values of the amplitude are zero.
Through the local analysis, the fundamental rogue wave for short wave can be classified into three patterns.
Without loss of generality, we take $\rho=1$, then there are two different cases:

In the case of $\sigma=1$:
(a) Dark state ($|\alpha|\leq \frac{\sqrt{-378+66\sqrt{33}}}{12}$): two local maxima at $(x_3,t_3)$ and one local minimum at $(x_1,t_1)$.
Especially, when $|\alpha|= \frac{\sqrt{-378+66\sqrt{33}}}{12}$, the local minimum at the characteristic point $(x_1,t_1)=(x_2,t_2)$.
(b) Intermediate state $(\frac{\sqrt{-378+66\sqrt{33}}}{12}<|\alpha| <\frac{\sqrt{-18+6\sqrt{33}}}{4})$:
two local maxima at $(x_3,t_3)$ and two local minima at the characteristic point $(x_2,t_2)$.
(c) Bright state ($ \frac{\sqrt{-18+6\sqrt{33}}}{4})\leq |\alpha|$ ):
one local maximum at $(x_1,t_1)$ and two local minima at $(x_2,t_2)$.
When the sign takes "=", the local maximum at $(x_1,t_1)=(x_3,t_3)$.

In the case of $\sigma=-1$ ($|\alpha|<\frac{3}{2}\sqrt{3}$ and $|\alpha|\neq\frac{\sqrt{2}}{2}$):
(a) Dark state ($\frac{\sqrt{378+66\sqrt{33}}}{12} \leq |\alpha|< \frac{3}{2}\sqrt{3}$): two local maxima at $(x_3,t_3)$ and one local minimum at $(x_1,t_1)$.
Especially, when $|\alpha|= \frac{\sqrt{378+66\sqrt{33}}}{12}$, the local minimum at the characteristic point $(x_1,t_1)=(x_2,t_2)$. (b) Intermediate state $(\frac{\sqrt{18+6\sqrt{33}}}{4} <|\alpha| <\frac{\sqrt{378+66\sqrt{33}}}{12})$:
two local maxima at $(x_3,t_3)$ and two local minima at the characteristic point $(x_2,t_2)$.
(c) Bright state ( $  |\alpha| \leq \frac{\sqrt{18+6\sqrt{33}}}{4}$ ):
(i) $  \frac{\sqrt{2}}{2} <|\alpha| \leq \frac{\sqrt{18+6\sqrt{33}}}{4}$:
one local maximum at $(x_1,t_1)$ and two local minima at $(x_2,t_2)$;
(ii) $  |\alpha| < \frac{\sqrt{2}}{2} $:
one local maximum at $(x_1,t_1)$ and two local minima at $(x_3,t_3)$.
When the sign takes "=", the local maximum at $(x_1,t_1)=(x_3,t_3)$.

For the LW component $L$, it possesses extrema $(x_1,t_1)$ and
\begin{eqnarray}
&& (x_4,t_4)=\left(\frac{2(\zeta^2_1+3\zeta^2_2)}{\zeta_1}\mu_3-\frac{1}{2\zeta_1}, \frac{\zeta_2}{\zeta_1}\mu_3 \right),
\end{eqnarray}
with $\mu_3=\frac{1}{4(\zeta^2_1+\zeta^2_2)} \pm \frac{\sqrt{3}}{4} \frac{\sqrt{\zeta^2_2(\zeta^2_1+4\zeta^2_2)}}{(\zeta^2_1+\zeta^2_2)(\zeta^2_1+4\zeta^2_2)}$.
The further local analysis shows that in both cases $\sigma=\pm 1$, the rogue wave for the LW component only exhibits bright state with one local maximum at $(x_1,t_1)$ and two local minima at $(x_4,t_4)$.

\begin{figure}[!htbp]
\centering
\includegraphics[height=1.8in,width=5.5in]{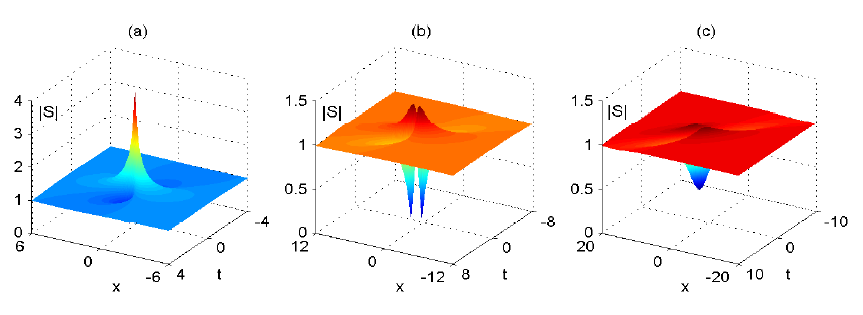}
\caption{First-order rogue wave for the SW in the LWSW model (\ref{dyo-rw-1})-(\ref{dyo-rw-2}) with the parameters $\rho=-\sigma=1$ and (a) bright state $\alpha=1.5$; (b) intermediate state $\alpha=2.1$ and (c) dark state $\alpha=2.3$.\label{first-oder-S}}
\end{figure}

\begin{figure}[!htbp]
\centering
\includegraphics[height=1.8in,width=5.5in]{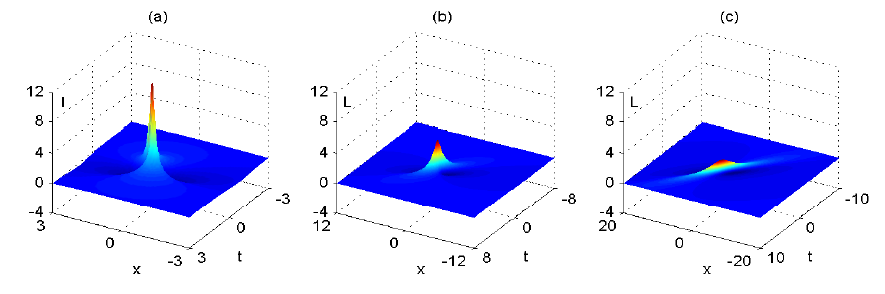}
\caption{First-order rogue wave for the LW in the LWSW model (\ref{dyo-rw-1})-(\ref{dyo-rw-2}) with the parameters $\rho=-\sigma=1$ and (a) $\alpha=1.5$; (b) $\alpha=2.1$ and (c) $\alpha=2.3$, respectively.\label{first-oder-L}}
\end{figure}

As illustrated in Fig. \ref{first-oder-S}, three types of RW exhibit different local structures for the SW when the parameter $\alpha$ takes
the value in its corresponding regimes.
It implies that RW's pattern for the SW component is dependent on $\alpha$, which decides the number, the position, the height and the type of extrema.
Fig \ref{first-oder-L} displays the RW states of the LW, three cases correspond to the same parameters' choices as shown in Fig.\ref{first-oder-S}.
It is easy to see that as $\alpha$ increases, the local structure of the LW always features the bright RW only the central amplitude decreases.

For the LWSW model (\ref{dyo-rw-1})-(\ref{dyo-rw-2}), it can be viewed as a coupling extension of the derivative NLS (DNLS) equation, exactly the Chen-Lee-Liu (CLL) equation.
We recall that the DNLS equation only supports the RW of bright type with two zero-amplitude points.
Here, due to the coupling component $L$ which leads to complex interplay between the dispersion
and the nonlinearity, dark and intermediate RWs appear for the SW component.
From the local analysis, one can know that two kinds of the bright RW emerge under the different parameters' conditions.
Specifically, the normal bright RW with two characteristic points can be realized in the regions  $|\alpha|\geq\frac{\sqrt{-18+6\sqrt{33}}}{4}$ ($\sigma=1$) and $\frac{\sqrt{2}}{2} <|\alpha| \leq \frac{\sqrt{18+6\sqrt{33}}}{4}$ ($\sigma=-1$).
The another case ($|\alpha| < \frac{\sqrt{2}}{2}$ for $\sigma=-1$) corresponds to a sepcial bright RW which also possesses one maximum and two minima amplitudes but two local minima are not characteristic points.
The intermediate RW is always characterized by two local maxima and two local minima at zero-amplitude points.
For the dark RW, its amplitude possesses two local maxima and one local minimum which usually can't attain zero.
But at the critical cases between the dark RW and the intermediate one in which $\alpha$ takes $|\alpha|= \frac{\sqrt{-\sigma 378+66\sqrt{33}}}{12}$, the local minimum occurs at zero-amplitude points.
In this situation, the dark RW reduces to a special one which can be referred to a black RW.

\subsection{High-order rogue wave}

The second-order rogue wave solution is obtained from Eqs.(\ref{theoremeq01})--(\ref{theoremeq04}) with $N=2$.
In this case, setting $a^{(0)}_0=b^{(0)}_0=1$, $a^{(0)}_1=b^{(0)}_1=0$, $a^{(0)}_2=b^{(0)}_2=0$, we obtain
the functions $f$ and $g$ as follows
\begin{eqnarray}\label{second-solution}
f
=\begin{vmatrix}
m_{11}^{(1,1,-1,0)} & m_{13}^{(1,0,-1,0)}  \\
m_{31}^{(0,1,-1,0)} & m_{33}^{(0,0,-1,0)}
\end{vmatrix},\ \
g
=\begin{vmatrix}
m_{11}^{(1,1,-1,1)} & m_{13}^{(1,0,-1,1)}  \\
m_{31}^{(0,1,-1,1)} & m_{33}^{(0,0,-1,1)}
\end{vmatrix}\,,
\end{eqnarray}
where the elements are determined by
\begin{eqnarray*}
&& m_{11}^{(1,1,n,k)} = A^{(1)}_1 B^{(1)}_1
\left.m^{(n,k)} \right|_{p=\zeta,q=\zeta^*},\ \
 m_{13}^{(1,0,n,k)} = A^{(1)}_1 B^{(0)}_3
\left.m^{(n,k)} \right|_{p=\zeta,q=\zeta^*},\\
&& m_{31}^{(0,1,n,k)} = A^{(0)}_3 B^{(1)}_1
\left.m^{(n,k)} \right|_{p=\zeta,q=\zeta^*},\ \
 m_{33}^{(0,0,n,k)} = A^{(0)}_3 B^{(0)}_3
\left.m^{(n,k)} \right|_{p=\zeta,q=\zeta^*},
\end{eqnarray*}
and the differential operators $A^{(1)}_1 = a^{(1)}_0 (p- {\rm i}\alpha) \partial_p +a^{(1)}_1$,
$B^{(1)}_1 = a^{(1)*}_0 (q+ {\rm i}\alpha) \partial_q +a^{(1)*}_1$,
$A^{(0)}_3 = \frac{1}{6} [(p- {\rm i}\alpha) \partial_p]^3 +a^{(0)}_3$,
$B^{(0)}_3 = \frac{1}{6} [(q+ {\rm i}\alpha) \partial_q]^3 +a^{(0)*}_3$
with
$a^{(1)}_{0}= 2 (p-{\rm i}\alpha)^2 +  \frac{{\rm i}\sigma \alpha \rho^2}{p-{\rm i}\alpha} +{\rm i}\alpha(p-{\rm i}\alpha)$ and
$a^{(1)}_{1}= \frac{1}{3} \left[ 4 (p-{\rm i}\alpha)^2 -  \frac{{\rm i}\sigma\alpha\rho^2}{p-{\rm i}\alpha} + {\rm i}\alpha(p-{\rm i}\alpha) \right]$.

\begin{figure}[!htbp]
\centering
\includegraphics[height=1.7in,width=5.5in]{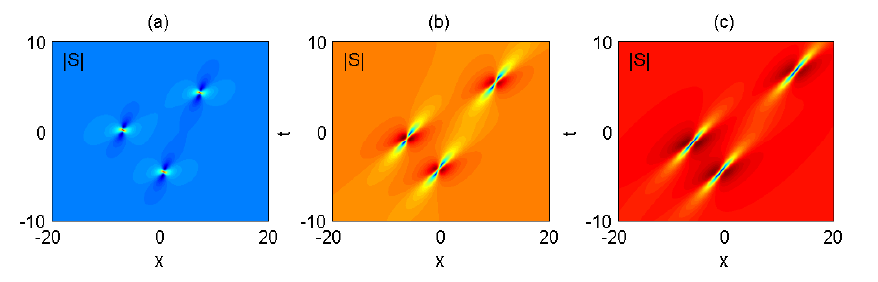}
\caption{Second-order rogue wave for the SW in the LWSW model (\ref{dyo-rw-1})-(\ref{dyo-rw-2}) with the parameters $\rho=-\sigma=1$, $a_0^{(0)}=1,a^{(0)}_1=a^{(0)}_2=0,a^{(0)}_3=350$ and (a) bright state $\alpha=1.5$; (b) intermediate state $\alpha=2.1$ and (c) dark state $\alpha=2.3$. \label{seconder-oder}}
\end{figure}

\begin{figure}[!htbp]
\centering
\includegraphics[height=1.7in,width=5.5in]{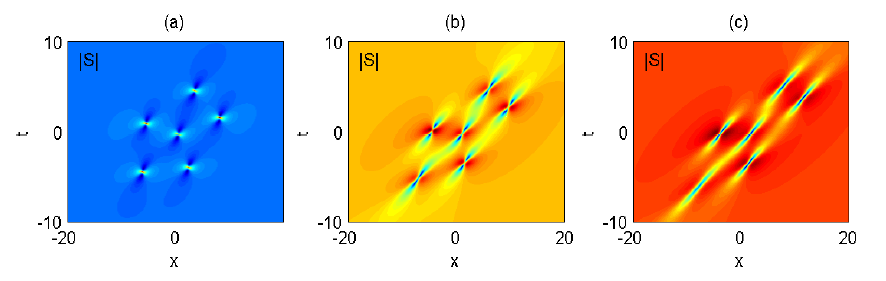}
\caption{Third-order rogue wave for the SW in the LWSW model (\ref{dyo-rw-1})-(\ref{dyo-rw-2}) with the parameters $\rho=-\sigma=1$, $a_0^{(0)}=1,a^{(0)}_1=a^{(0)}_2=a^{(0)}_3=a^{(0)}_4=0,a^{(0)}_5=2000$ and (a) bright state $\alpha=1.5$; (b) intermediate state $\alpha=2.1$ and (c) dark state $\alpha=2.3$.\label{third-oder}}
\end{figure}

Since the LW always features a bright RW, we only present the configuration of the SW to illustrate higher RWs.
Three second-order RWs for the SW are displayed in Fig.\ref{seconder-oder}, each one takes the same value of the parameter $\alpha$ as one
shown in Fig.\ref{first-oder-S}.
One can observe that second-order RWs manifest the superposition of three fundamental ones and they obey the triangle arrays.
Owing to the same parameters $\alpha$ as the first-order cases respectively, three second-order RWs exhibit pure dark, intermediate and bright RW's combinations individually.

For third and higher-order RWs, which describe the superposition of more fundamental RWs, one need to take larger $N$ in (\ref{theoremeq01})-(\ref{theoremeq06}). The expressions are too complicated to be written here.
Fig.\ref{third-oder} shows the third-order RW for $N=3$ graphically, in which three plots still take the same parameter $\alpha$ as Figs.\ref{first-oder-S} and \ref{seconder-oder}.
It can be seen that third-order RWs exhibit the superposition of six fundamental ones and they constitute a shape of pentagon.
Besides, each combination only contains one type of elementary RW purely in three cases, which coincides with ones of first and second-order cases.
This fact suggests that only three types of RW happens whether the RW is fundamental or higher-order one for the SW and the RW's pattern completely depends on the parameter $\alpha$.
For instance, when $\sigma=1$, three types of RW (fundamental one and higher-order superposition) strictly takes place at
at three intervals of $\alpha$, i.e. $|\alpha|\leq \frac{\sqrt{-378+66\sqrt{33}}}{12}$ for dark state, $\frac{\sqrt{-378+66\sqrt{33}}}{12}<|\alpha| <\frac{\sqrt{-18+6\sqrt{33}}}{4}$ for
intermediate state and $ \frac{\sqrt{-18+6\sqrt{33}}}{4})\leq |\alpha|$ for bright state.
That is to say, the construction of higher-order RW solutions here don't allow the mixed superposition among different
types of fundamental RWs.

\section{Modulation Instability}

Next we investigate the linear stability analysis of the LWSW model by considering small perturbations of the form
$S= [a+ S_1]e^{{\rm i}[\omega x-(\omega^2-2b\omega) t]}$ and $L=b+L_1$.
The substitution yields a group of linearized partial differential equations.
Recalling that $L$ is real, we can assume the perturbations are space periodic with the fixed frequency $\Omega$, i.e.,
$S_1=s_1(t)e^{{\rm i}\Omega x}+s_2(t)e^{-{\rm i}\Omega x}$ and
$L_1=l(t)e^{{\rm i}\Omega x} +l^*(t)e^{-{\rm i}\Omega x}$,
which leads the above linearized partial differential equations into a group of linear ordinary differential equations
$s'={\rm i}Ms$ with $s=[s_1,s^*_2,l]^T$ and
\begin{eqnarray}
M=
\left[
\begin{matrix}
-\Omega^2+2b\Omega-2\omega\Omega & 0 & 2a\omega  \\
0 & \Omega^2+2b\Omega-2\omega\Omega & -2a\omega  \\
-2a\sigma\Omega & -2a\sigma\Omega & 0
\end{matrix}
\right].
\end{eqnarray}
This set of differential equations with the real frequency $\Omega$ suggests that the functions
$s_1(t)$, $s_2(t)$ and $l(t)$ are the linear combinations of exponentials $\exp({\rm i}\lambda_jt)$
where $\lambda_j$, $j=1,2,3$ represent three eigenvalues of the matrix $M$.
Such eigenvalues are given by the roots of the characteristic polynomial $P(\lambda)$ of the matrix $M$,
\begin{eqnarray}
&& P(\lambda)= \lambda^3 +P_2\lambda^2+P_1\lambda +P_0,
\end{eqnarray}
with $P_2=4(\omega-b)\Omega$, $P_1=[4(b-\omega)^2-\Omega^2]\Omega^2$ and $P_0=-8\sigma \omega a^2 \Omega^3$.

\begin{figure}[!htbp]
\centering
\includegraphics[height=1.8in,width=4.4in]{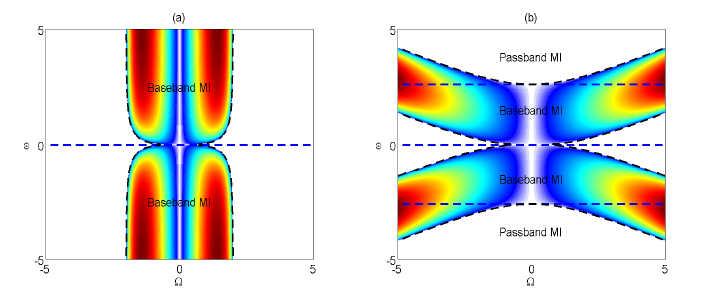}
\caption{ MI for the LWSW model (\ref{dyo-rw-1})-(\ref{dyo-rw-2}) on the $(\Omega,\omega)$ plane, calculated with the parameters $a=1$, $b=0$, (a):$\sigma=1$ and (b):$\sigma=-1$. Dark dashed curves represent the analytical marginal stability condition $\Delta=0$. \label{mi-figs}}
\end{figure}

It is known that when an eigenvalue has a negative imaginary part, MI will occur with the exponential growing perturbation.
From the matrix $M$, one can find each entry is real, so the corresponding eigenvalues are either real root, or a pair of complex-conjugate roots.
More specifically, we calculate the discriminant of the characteristic polynomial $P(\lambda)$ as
\begin{eqnarray}
\Delta=4\Omega^6\left\{
3\Omega^6
-24(b-\omega)^2\Omega^4
-48(b-\omega)[9\sigma \omega a^2-(b-\omega)^3]\Omega^2
-48\sigma\omega a^2[27\sigma \omega a^2-4(b-\omega)^3]
\right\}.
\end{eqnarray}
Then, $\Delta>0$ results in real roots for the polynomial $P(\lambda)$, which implies that no MI appears,
whereas $\Delta<0$ yields two complex conjugate roots, which suggests that MI exists.
The marginal stability curves corresponds to the discriminant $\Delta=0$.
Without loss of generality, by taking $a=1$ and $b=0$, MI gain spectrums of the LWSW model (\ref{dyo-rw-1})-(\ref{dyo-rw-2})
are displayed for two kinds of nonlinearity $\sigma=1$ and $\sigma=-1$ in Fig.\ref{mi-figs} respectively.

As analyzed in \cite{baronio2014vector,baronio2015baseband},
baseband MI defined as the condition where an unstable continuous-wave background corresponds to perturbations infinitesimally small frequencies, is responsible for RW excitation, whereas passband MI which means the perturbation undergoes
gain in a spectral region not including zero frequency as a limiting case, doesn't support the RW.
Thus the limit situation where $\Omega \rightarrow 0$ decides the occurrence of baseband MI.
In this case, the discriminant of the polynomial $P(\lambda)$ degenerates to $\Delta=-48\sigma\omega (4\omega^3+27\sigma\omega)$,
which gives rise to two cases:
(1) $\sigma=1$, no MI and (2) $\sigma=-1$, MI condition $|\omega|<\frac{3}{2}\sqrt{3}$.
The coincidence is that the baseband MI condition is exactly one for the existence of rogue-wave solutions.

\section{Conclusion}

The long wave-short wave model describes the interaction between the long wave and the short wave.
Exact higher-order rational solution expressed by determinants is calculated via the Hirota's bilinear method and the KP hierarchy reduction.
It is found that the fundamental rogue wave for the SW can be classified into three different patterns: bright, intermediate and
dark ones, whereas the rogue wave for the LW is always bright type.
The higher-order rogue waves correspond to the superposition of fundamental rogue waves.
The modulation instability  analysis show that the condition of the baseband modulation instability where an unstable continuous-wave background corresponds to perturbations with infinitesimally small frequencies, coincides with the condition for the existence of rogue-wave solutions.

\section*{Acknowledgments}
J.C. acknowledges support from National
Natural Science Foundation of China (No.11705077). B.F.F. was partially supported by NSF Grant under Nos. 171599 and the COS Research Enhancement Seed Grants Program at UTRGV.
KM's work was supported by
JSPS Grant-in-Aid for Scientific Research (C-15K04909) and JST CREST.

\section*{Appendix}
\setcounter{equation}{0}
 \renewcommand\theequation{A\arabic{equation}}

In this Appendix we will prove Theorem 2.1 in Sec. II via the KP hierarchy reduction.
First we present the following lemma.

\emph{Lemma 1} The bilinear equations in the extended KP hierarchy
\begin{eqnarray}
\label{dark-kp-01}&& (D_{x_2} - 2a D_{x_1} - D^2_{x_1})\tau_{n,k+1}\cdot \tau_{n,k}=0,\\
\label{dark-kp-02}&& (D_{x_2} + D^2_{x_1})\tau_{n,k}\cdot \tau_{n+1,k}=0,\\
\label{dark-kp-03}&& (aD_{t_a}+1)\tau_{n,k}\cdot \tau_{n+1,k}=\tau_{n,k+1}\tau_{n+1,k-1},
\end{eqnarray}
where $a$ is a complex constant, $n$ and $k$ are
integers, have the following solution
\begin{eqnarray}
\tau_{n,k}
=\tilde{m}^{(n,k)}
= \frac{{\rm i}p}{p+q}
\left(-\frac{p}{q}\right)^n
\left(-\frac{p-a}{q+a} \right)^k
{\rm e}^{\tilde{\xi}+\tilde{\eta}},
\end{eqnarray}
with
\begin{eqnarray*}
\tilde{\xi}= \frac{1}{p-a} t_a+ p x_1 +p^2 x_2,\ \
\tilde{\eta}= \frac{1}{q+a} t_a+ qx_1 - q^2x_2,
\end{eqnarray*}
where $p$, $q$ and $a$ are complex parameters.

In order to derive the rational soluiton, we introduce the differential operators $A_k^{(\nu)}$ and $B_l^{(\mu)}$  with respect to $p$ and $q$ respectively
\begin{eqnarray}
&& A_n^{(\nu)}=
\sum_{k=0}^na_k^{(\nu)}\frac{[(p-a)\partial_p]^{n-k}}{(n-k)!},\quad
 n \geq 0,
\\
&& B_n^{(\nu)}=
\sum_{k=0}^nb_k^{(\nu)}\frac{[(q+a)\partial_q]^{n-k}}{(n-k)!},\quad
 n\geq0,
\end{eqnarray}
where $a_k^{(\nu)}$ and $b_k^{(\nu)}$ are constants satisfying the iterated relations
\begin{eqnarray}
&& a^{(\nu+1)}_{k}=\sum^{k}_{j=0} \frac{2^{j+2} (p-a)^2 + (-1)^{j}  \frac{\lambda a}{p-a} +2a(p-a)}{ (j+2)!} a^{(\nu)}_{k-j},
\ \ \nu=0,1,2,\cdots,\\
&& b^{(\nu+1)}_{k}=\sum^{k}_{j=0} \frac{2^{j+2} (q+a)^2 - (-1)^{j}  \frac{\lambda a}{q+a} -2a(q+a)}{ (j+2)!} b^{(\nu)}_{k-j},
\ \ \nu=0,1,2,\cdots\,.
\end{eqnarray}

Based on the Leibniz rule, one has
\begin{eqnarray}
\nonumber  [(p-a)\partial_p]^m\left(p^2+ \frac{\lambda a}{p-a}\right) &=& \sum^m_{l=0}\left( \begin{array}{c} m \\ l \end{array} \right) \left[2^l (p-a)^2 + (-1)^l\frac{\lambda a}{p-a} +2a(p-a) \right][(p-a)\partial_p]^{m-l} \\
 && + a^2[(p-a)\partial_p]^{m},
\end{eqnarray}
and
\begin{eqnarray}
\nonumber  [(q+a)\partial_q]^m\left(q^2- \frac{\lambda a}{q+a}\right) &=& \sum^m_{l=0}\left( \begin{array}{c} m \\ l \end{array} \right) \left[2^l (q+a)^2 - (-1)^l\frac{\lambda a}{q+a} - 2a(q+a) \right][(q+a)\partial_q]^{m-l} \\
 && + a^2[(q+a)\partial_q]^{m}\,.
\end{eqnarray}

Furthermore, we can derive
\begin{eqnarray*}
&& \left[A^{(\nu)}_n,p^2+\frac{\lambda a}{p-a}\right]
=\sum_{k=0}^{n-1}\frac{a_k^{(\nu)}}{(n-k)!}
\left[((p-a)\partial_p)^{n-k},p^2+\frac{\lambda a}{p-a}\right] \\
&& =\sum_{k=0}^{n-1}\frac{a_k^{(\nu)}}{(n-k)!}
\sum^{n-k}_{l=1}\left( \begin{array}{c} n-k \\ l \end{array} \right)
\left(2^l (p-a)^2 + (-1)^l\frac{\lambda a}{p-a} +2a(p-a) \right)
((p-a)\partial_p)^{n-k-l},
\end{eqnarray*}
where $[\ ,\ ]$ is the commutator defined by $[X,Y]=XY-YX$.

Let $\tilde{\zeta}$ be the solution of the algebraic equation
$$
2p- \frac{\lambda a}{(p-a)^2}=0.
$$
Hence we have
$$
\left.\left[A^{(\nu)}_n,p^2+\frac{\lambda a}{p-a}\right]\right|_{p=\tilde{\zeta}}=0,
$$
for $n=0,1$ and
\begin{eqnarray*}
&&\left.\left[A^{(\nu)}_n,p^2+\frac{\lambda a}{p-a}\right]\right|_{p=\tilde{\zeta}} \\
&&=\left.\sum_{k=0}^{n-2}\frac{a_k^{(\nu)}}{(n-k)!}
\sum^{n-k}_{l=2}\left( \begin{array}{c} n-k \\ l \end{array} \right)
\left\{2^l (p-a)^2 + (-1)^l\frac{\lambda a}{p-a} +2a(p-a) \right\}
[(p-a)\partial_p]^{n-k-l}\right|_{p=\tilde{\zeta}} \\
&&=\left.\sum_{k=0}^{n-2}\sum^{n-k-2}_{j=0}
\frac{a^{(\nu)}_k}{(j+2)!(n-k-j-2)!}
\left\{2^{j+2} (p-a)^2 + (-1)^{j} \frac{\lambda a}{p-a} +2a(p-a) \right\}
[(p-a)\partial_p]^{n-k-j-2}\right|_{p=\tilde{\zeta}} \\
&&=\left.\sum^{n-2}_{\hat{k}=0} \left( \sum^{\hat{k}}_{\hat{j}=0} \frac{2^{\hat{j}+2} (p-a)^2 + (-1)^{\hat{j}}  \frac{\lambda a}{p-a} +2a(p-a)}{ (\hat{j}+2)!} a^{(\nu)}_{\hat{k}-\hat{j}} \right) \frac{[(p-a)\partial_p]^{n-2-\hat{k}} }{(n-2-\hat{k})!}\right|_{p=\tilde{\zeta}} \\
&&=\left.\sum^{n-2}_{\hat{k}=0} a^{(\nu+1)}_{\hat{k}} \frac{[(p-a)\partial_p]^{n-2-\hat{k}} }{(n-2-\hat{k})!}\right|_{p=\tilde{\zeta}} \\
&&=\left.  A^{(\nu+1)}_{n-2}\right|_{p=\tilde{\zeta}},
\end{eqnarray*}
for $n\ge2$.
Thus the differential operator $A_n^{(\nu)}$ satisfies the following relation
\begin{eqnarray}
&&\left. \left[A_n^{(\nu)},p^2+\frac{2{\rm i}}{p-a}\right]\right|_{p=\tilde{\zeta}}
=\left.A_{n-2}^{(\nu+1)}\right|_{p=\tilde{\zeta}},
\end{eqnarray}
where we define $A_n^{(\nu)}=0$ for $n<0$.

Similarly, it is shown that the differential operator $B_n^{(\nu)}$ satisfies
\begin{eqnarray}
&&\left. \left[B_n^{(\nu)},q^2-\frac{\lambda a}{q+a}\right]\right|_{q=\bar{\tilde{\zeta}}}
=\left.B_{n-2}^{(\nu+1)}\right|_{q=\bar{\tilde{\zeta}}},
\end{eqnarray}
where we define $B_n^{(\nu)}=0$ for $n<0$ and $\bar{\tilde{\zeta}}$ needs to satisfy
\begin{eqnarray}
2q+\frac{\lambda a}{(q+a)^2}=0.
\end{eqnarray}

Consequently, by referring to above two relations,  we have
\begin{eqnarray*}
&& \left. (\partial_{x_2}+\lambda a\partial_{t_a})\tilde{m}_{ls}^{(\nu,\mu, n,k)}\right|_{p=\tilde{\zeta},q=\bar{\tilde{\zeta}}}
\\
&=& \left. \left[A_l^{(\nu)}B_s^{(\mu)}
(\partial_{x_2}+\lambda a\partial_{t_a})\tilde{m}^{(n,k)}\right]\right|_{p=\tilde{\zeta},q=\bar{\tilde{\zeta}}}
\\
&=& \left.\left(A_l^{(\nu)}B_s^{(\mu)}
\left(p^2-q^2+\lambda a \left(\frac{1}{p-a}+\frac{1}{q+a}\right)\right) \tilde{m}^{(n)}\right)\right|_{p=\tilde{\zeta},q=\bar{\tilde{\zeta}}}
\\
&=& \left.\left(A_l^{(\nu)}\left(p^2+\frac{\lambda a }{p-a}\right)B_s^{(\mu)}
\tilde{m}^{(n)}\right)\right|_{p=\tilde{\zeta},q=\bar{\tilde{\zeta}}}
-\left.\left(A_l^{(\nu)}B_s^{(\mu)}\left(q^2-\frac{\lambda a }{q+a}\right)
\tilde{m}^{(n)}\right)\right|_{p=\tilde{\zeta},q=\bar{\tilde{\zeta}}}
\\
&=& \left.\left\{\left(\left(p^2+\frac{\lambda a}{p-a}\right)A_l{(\nu)}+A_{l-2}^{(\nu+1)}\right)B_s^{(\mu)}
\tilde{m}^{(n,k)}\right\}\right|_{p=\tilde{\zeta},q=\bar{\tilde{\zeta}}}
\\
&&
-\left.\left\{A_l^{(\nu)}\left(\left(q^2-\frac{\lambda a}{q+a}\right)B_s^{(\mu)} + B_{s-2}^{(\mu+1)}\right)
\tilde{m}^{(n,k)}\right\}\right|_{p=\tilde{\zeta},q=\bar{\tilde{\zeta}}}
\\
&=& \left(\tilde{\zeta}^2+\frac{\lambda a}{\tilde{\zeta}-a}\right) \left.\tilde{m}_{ls}^{(\nu,\mu, n,k)}\right|_{p=\tilde{\zeta},q=\bar{\tilde{\zeta}}}
+\left.\tilde{m}_{l-2,s}^{(\nu+1,\mu, n,k)}\right|_{p=\tilde{\zeta},q=\bar{\tilde{\zeta}}}
\\&&-\left(\bar{\tilde{\zeta}}^2-\frac{\lambda a}{\bar{\tilde{\zeta}}+a}\right) \left.\tilde{m}_{ls}^{(\nu,\mu, n,k)}\right|_{p=\tilde{\zeta},q=\bar{\tilde{\zeta}}}
-\left.\tilde{m}_{l,s-2}^{(\nu,\mu+1,n,k)}\right|_{p=\tilde{\zeta},q=\bar{\tilde{\zeta}}} .
\end{eqnarray*}

Then the differential of the following determinant
\begin{eqnarray*}
{\tilde{\tau}}_{n,k}=\det_{1\leq i,j\leq N}\left(\left.\tilde{m}_{2i-1,2j-1}^{(N-i,N-j, n,k)}\right|_{p=\tilde{\zeta},q=\bar{\tilde{\zeta}}}\right)
\end{eqnarray*}
can be calculated as
\begin{eqnarray*}
&&(\partial_{x_2}+\lambda a \partial_{t_a}){\tilde{\tau}}_n \\
&=& \sum^N_{i=1} \sum^N_{j=1} \Delta_{ij} (\partial_{x_2}+\lambda a \partial_{t_a})\left( \left. \tilde{m}^{(N-i,N-j,n,k)}_{2i-1,2j-1}\right|_{p=\tilde{\zeta},q=\bar{\tilde{\zeta}}} \right)\\
&=& \sum^N_{i=1} \sum^N_{j=1} \Delta_{ij}
\left[\left(\tilde{\zeta}^2+\frac{\lambda a}{\tilde{\zeta}-a}\right)  \left. \tilde{m}^{(N-i,N-j,n,k)}_{2i-1,2j-1}\right|_{p=\tilde{\zeta},q=\bar{\tilde{\zeta}}} + \left. \tilde{m}^{(N-i+1,N-j,n,k)}_{2i-3,2j-1}\right|_{p=\tilde{\zeta},q=\bar{\tilde{\zeta}}}  \right.
\\
&& \left.
- \left(\bar{\tilde{\zeta}}^2-\frac{\lambda a}{\bar{\tilde{\zeta}}+a}\right)  \left. \tilde{m}^{(N-i,N-j,n,k)}_{2i-1,2j-1}\right|_{p=\tilde{\zeta},q=\bar{\tilde{\zeta}}}
- \left. \tilde{m}^{(N-i,N-j+1,n,k)}_{2i-1,2j-3}\right|_{p=\tilde{\zeta},q=\bar{\tilde{\zeta}}}  \right]\\
&=& \left(\tilde{\zeta}^2+\frac{\lambda a}{\tilde{\zeta}-a}\right) N {\tilde{\tau}}_n
+  \sum^N_{i=1} \sum^N_{j=1} \Delta_{ij} \left. \tilde{m}^{(N-i+1,N-j,n,k)}_{2i-3,2j-1}\right|_{p=\tilde{\zeta},q=\bar{\tilde{\zeta}}}
\\&& - \left(\bar{\tilde{\zeta}}^2-\frac{\lambda a}{\bar{\tilde{\zeta}}+a}\right) N {\tilde{\tau}}_n
 -  \sum^N_{i=1} \sum^N_{j=1} \Delta_{ij} \left. \tilde{m}^{(N-i,N-j+1,n,k)}_{2i-1,2j-3}\right|_{p=\tilde{\zeta},q=\bar{\tilde{\zeta}}},
\end{eqnarray*}
where $\Delta_{ij}$ is the $(i,j)$-cofactor of the matrix $\left(\left.\tilde{m}_{2i-1,2j-1}^{(N-i,N-j, n)}\right|_{p=\tilde{\zeta},q=\bar{\tilde{\zeta}}} \right)_{1\leq i,j\leq N}$.
For the term $\sum^N_{i=1} \sum^N_{j=1} \Delta_{ij} \left. \tilde{m}^{(N-i+1,N-j,n)}_{2i-3,2j-1}\right|_{p=\tilde{\zeta},q=\bar{\tilde{\zeta}}}$,
it vanishes since for $i=1$ this summation is a determinant with the elements in first row being zero
and for $i=2,3,\ldots$ this summation is a determinant with two identical rows.
Similarly, the term $\sum^N_{i=1} \sum^N_{j=1} \Delta_{ij} \left. \tilde{m}^{(N-i,N-j+1,n)}_{2i-1,2j-3}\right|_{p=\tilde{\zeta},q=\bar{\tilde{\zeta}}}$ vanishes.
Therefore, ${\tilde{\tau}}_n$ satisfies the reduction condition
\begin{eqnarray}\label{red-condition}
(\partial_{x_2}+\lambda a \partial_{t_a}){\tilde{\tau}}_n = \left(\tilde{\zeta}^2-\bar{\tilde{\zeta}}^2+\frac{\lambda a}{\tilde{\zeta}-a }+\frac{\lambda a}{\bar{\tilde{\zeta}}+a}\right) N {\tilde{\tau}}_n,
\end{eqnarray}
such that these algebraic solutions ${\tilde{\tau}}_{n,k}$ satisfy the (1+1)-dimensional bilinear equations:
\begin{eqnarray}
\label{dark-kp-011}&& (D_{x_2} - 2a D_{x_1} - D^2_{x_1})\tilde{\tau}_{n,k+1}\cdot \tilde{\tau}_{n,k}=0,\\
\label{dark-kp-021}&& (D_{x_2} + D^2_{x_1})\tilde{\tau}_{n,k}\cdot \tilde{\tau}_{n+1,k}=0,\\
\label{dark-kp-031}&& D_{x_2}\tilde{\tau}_{n,k}\cdot \tilde{\tau}_{n+1,k}= \lambda(\tilde{\tau}_{n,k} \tilde{\tau}_{n+1,k}-\tilde{\tau}_{n,k+1}\tilde{\tau}_{n+1,k-1}).
\end{eqnarray}

By taking $a={\rm i}\alpha$, $\lambda=2\sigma\rho^2$, $\bar{\tilde{\zeta}}=\tilde{\zeta}^*$ and $x_1=x$, $x_2={\rm i}t$,
one can define
\begin{eqnarray}
f=\tau_{-1,0},\ \
g=\tau_{-1,1},\ \
f^*=\tau_{0,0},\ \
g^*=\tau_{0,-1}.
\end{eqnarray}
which reduce (\ref{dark-kp-011})-(\ref{dark-kp-031}) to the bilinear equations (\ref{bilinear-01})-(\ref{bilinear-03}).
From the reduction condition (\ref{red-condition}), $t_a$ becomes a dummy variable which can be taken as zero.
Therefore we can arrive at the Theorem 2.1.


\end{document}